\documentclass[useAMS,usegraphicx, usenatbib]{mn2e}

\usepackage{amsmath}
\usepackage{amssymb}

\newcommand{\msun}{\,\text{M}_\odot}
\newcommand{\nG}{\,\text{nG}}

\newcommand{\HII}{H\textsc{ii} }

\begin{document}

\title{Suppression of small baryonic structures  due to a primordial magnetic field}
\author[L. F. S. Rodrigues, R. S. de Souza and R. Opher]
{Luiz Felippe S. Rodrigues$^{1}$\thanks{Email: felippe@astro.iag.usp.br}, Rafael S. de Souza$^{1}$\thanks{Email: rafael@astro.iag.usp.br} and Reuven Opher$^{1}$\thanks{Email: opher@astro.iag.usp.br}
\\ $^{1}$IAG, Universidade de S\~{a}o Paulo, Rua do Mat\~{a}o 1226, Cidade
 Universit\'{a}ria, CEP 05508-900, S\~{a}o Paulo, SP, Brazil\\
}

\date{Accepted 2010 March 11.  Received 2010 March 11; in original form 2010 February 1}

\pagerange{\pageref{firstpage}--\pageref{lastpage}} \pubyear{2010}

\maketitle
\label{firstpage}

\begin{abstract}
We investigate the impact of the existence of a primordial magnetic field on the filter mass, characterizing the minimum baryonic mass that can form in dark matter (DM) haloes.
For masses below the filter mass, the  baryon content of DM haloes are severely depressed.
The filter mass is the mass when the baryon to DM mass ratio in a halo is equal to half the baryon to DM ratio of the Universe. 
The filter mass has previously been used in semianalytic calculations of galaxy formation, without taking into account the possible existence of a primordial magnetic field.
We examine here its effect on the filter mass.
For homogeneous comoving primordial magnetic fields of $B_0 \sim 1$ or 2 nG and a reionization epoch that starts at a redshift $z_s=11$ and is completed at $z_r=8$, the filter mass is increased at redshift 8, for example, by factors 4.1 and  19.8, respectively. The dependence of the filter mass on the parameters describing the reionization epoch is investigated.
Our results are particularly important for the formation of low mass galaxies in the presence of a homogeneous primordial magnetic field. For example, for $B_0\sim 1\nG$  and a reionization epoch of $z_s\sim 11$ and $z_r\sim7$, our results indicate that galaxies of total mass 
$M\sim5 \times 10^8\msun$
 need to form at redshifts 
$z_F\gtrsim 2.0$,
 and galaxies of total mass $M\sim10^8\msun$ at redshifts 
$z_F\gtrsim 7.7$.
\end{abstract}
 

\section{Introduction}
The formation of galaxies is one of the most important  areas in cosmology. Within the simplified hierarchical scenario of a Universe governed by a cosmological constant and cold dark matter (DM), the density profiles of forming DM haloes have been well characterized.
Numerical calculations show that the first generation of galaxies formed at very high redshifts inside collapsing haloes starting at $z \sim 65$, corresponding to high peaks of the primordial DM density field  \citep{NNB}. CMB observations  suggest that reionization began at high redshifts.
This means that a high abundance of luminous objects must have existed at that time, since these first luminous objects are expected to have heated and reionized their surroundings
\citep{rev,WL03,HH03,Cen}. 

The formation of a luminous object inside a halo necessarily  requires the existence of baryonic gas  inside the halo. Even in haloes that are too small for cooling via atomic hydrogen, the gas content can have substantial, and observable, astrophysical effects. In addition to the possibility of hosting astrophysical sources, such as stars, small haloes may produce a 21-cm signature \citep{Kuhlen,Shapiro06,NB08, Furlanetto06}, and can
block ionizing radiation and produce an overall delay in the global progress of reionization \citep{bl02, iliev2, iss05, mcquinn07}. 

As shown by \citet{Gnedin2000a} and \citet{Gnedin1997}, both in the linear and non-linear regimes, the accretion of gas into DM haloes is suppressed below a characteristic mass scale called the filter mass, $M_F$. This mass scale coincides with the Jeans mass, $M_J$, if the latter does not vary in time. Otherwise, $M_F$ is a time average of $M_J$. Thus, an increase in the ambient pressure  in the past, causes an increase in $M_J$ and suppresses the accretion of baryons into DM haloes in a cumulative fashion, producing an increase in $M_F$.

Until now, studies focused on the UV heating of the neutral interstellar gas as the main source of pressure, for determining  the  filter mass. These results  are widely used in many semianalytic models (e.g. \citealt{Maccio2009}), particularly those designed to study the properties of small galaxies (due to the high redshift character of the UV heating).

In this paper we add the effect of a possible homogeneous primordial magnetic field as another important source of ambient pressure. The magnetic field contributes to pressure support, which changes the Jeans mass and, consequently, the filter mass and the quantity of gas that is accreted by DM haloes.

The paper is organized as follows. In section \ref{sec:magnetic} we briefly review  our knowledge about primordial magnetic fields and in section \ref{sec:filter} we present our results, calculating the filter mass with a Jeans mass corrected for the presence of a primordial magnetic field. Finally, in section \ref{sec:conclusions}, we present our conclusions.

\section{Primordial Magnetic Fields} \label{sec:magnetic}

The origin of large-scale cosmic magnetic fields in galaxies and protogalaxies remains   a challenging problem  in astrophysics \citep{Zweibel1997, Kulsrud2008, Souza2008, Souza2009, Widrow2002}. Understanding the origin of the structures of the present Universe requires a knowledge of the origin of magnetic fields. The magnetic fields fill interstellar and intracluster space and affect the evolution of galaxies and galaxy clusters. 
There have been many attempts to explain  the origin of cosmic magnetic  fields.
One of the most popular astrophysical theories for creating seed primordial fields is that they were generated by the Biermann mechanism \citep{Biermann1950}. It has been suggested that this mechanism acts in diverse astrophysical systems, such as  large scale structure formation \citep{Peebles1967, Rees1972, Wasserman1978}, cluster of galaxies \citep{lagana},  cosmological ionizing fronts \citep{Gnedin2000a},  gamma ray bursts \citep{souza10},  star formation and  supernova explosions  \citep{Hanayama2005, Miranda1998}. Another mechanism for creating cosmic magnetic fields was suggested by  \citet{Ichiki2006}. They  investigated the second-order couplings between photons and electrons  as a possible origin of  magnetic fields on cosmological scales before the epoch of recombination.  Studies of magnetic field generation,   based
on cosmological perturbations, have also been made  \citep{Takahashi2005, Takahashi2006, Clarke2001, Maeda2009}. 

In our galaxy, the magnetic field is coherent over kpc scales with alternating directions 
in the arm and inter-arm regions (e.g.,\citealt{Kronberg1994, han08}). Such alternations are expected for magnetic fields of primordial origin \citep{Grasso2001}.
 
Various observations put upper limits on  the intensity of  a homogeneous primordial magnetic field. Observations of the small-scale cosmic microwave background (CMB) anisotropy yield an upper comoving limit of $4.7\nG$ for a homogeneous primordial field \citep{Yamazaki2006}. Reionization of the Universe puts upper limits of $0.7-3 \nG$ for a homogeneous primordial field, depending on the assumptions of the stellar population that is responsible for reionizing the Universe \citep{Schleicher2008}. 

De Souza \& Opher (2008) suggested that the fluctuations of the plasma predicted by the Fluctuation Dissipation Theorem after the quark-hadron transition (QHT) is a natural source for a present primordial magnetic field. They evolved the fluctuations after the QHT to the present era and predict a present cosmic inhomogeneous web of primordial magnetic fields and not a homogeneous one. The effect of this cosmic web of magnetic fields on galaxy formation will be investigated in a future article.

\section{Filter Mass} \label{sec:filter}

In linear theory the filter mass, $M_F$,  first defined by \cite{Gnedin1997}, describes the highest DM mass scale for which the baryon accretion is suppressed significantly. 
As shown by Gnedin \& Hui, there is a characteristic length scale, called the filtering scale, over which the baryonic perturbations are smoothed as compared to the dark matter ones.  The filtering scale is (characterized by the wavenumber $k_F$) defined as (see also, \citealt{Naoz2006}) 
\begin{equation}
 \frac{\delta_b}{\delta_{tot}}=1-\frac{k^2}{k_F^2} \,\text{ ,}
\end{equation}
where $\delta_b$ is the density contrast of baryonic matter and $\delta_{tot}$, the total density contrast. For $k$ comparable to $k_F$, the density contrast $\delta_b$ is severely depressed.

 Following \citet{Gnedin2000a}, we can relate the comoving wavenumber associated with this length scale with the Jeans wavenumber by the equation 
\begin{equation}
 \frac{1}{k^2_F(a)}=\frac{3}{a}\int^a_0 \frac{da'}{k_J^2(a')} \left[ 1-\left(\frac{a'}{a}\right)^\frac{1}{2} \right]\label{eq:kj}
\end{equation}
where a flat matter dominated universe is  assumed. In the case with no magnetic fields $k_{J}$ can be written as

\begin{equation}
k_{J} = \left(\frac{4\pi G \rho}{c_{s}^{2}}\right)^{1/2}, 
\label{eq:kJ}
\end{equation} 
where $c_{s}$ is the sound velocity of the medium. 
We see that, as mentioned above, the overall suppression of the growth of baryonic density perturbations depends on a time-average of the Jeans scale. By translating the length scales into mass scales, we can then define the Jeans mass and filter mass,
\begin{equation}
  M_J \equiv \frac{4\pi}{3} \bar{\rho} \left(\frac{2\pi a}{k_J}\right)^3\,\,\text{ and }\,\, M_F \equiv \frac{4\pi}{3} \bar{\rho} \left(\frac{2\pi a}{k_F}\right)^3 \,\text{ .}
  \label{eq:MJMF}
\end{equation}

From these definitions and equation \ref{eq:kj}, we can write,
\begin{equation}
 M_F^\frac{2}{3}=\frac{3}{a} \int^a_0 da'\,\, M_J^\frac{2}{3}(a')\left[ 1-\left(\frac{a'}{a}\right)^\frac{1}{2}\right]\text{ .}\label{eq:filter}
\end{equation}

A very important result, obtained by \cite{Gnedin2000a}, through the study of cosmological hydrodynamical simulations, is that the filter mass still characterizes the scale of suppression of the formation of baryonic structures even in the nonlinear regime. Namely, they found that the filter mass corresponds to the characteristic halo mass where the halo's baryonic gas fraction is approximately 50\% of the cosmic baryon fraction. 

A simple way to take into account the effects of the magnetic field on the filter mass, is the 
 generalization of equation (\ref{eq:MJMF}), replacing  equation (\ref{eq:kJ}) by 
\begin{equation}
k_{JB} = \left(\frac{4\pi G \rho}{c_{s}^{2}+v_{A}^{2}}\right)^{1/2},
\end{equation}
where $v_A$ is the Alfven velocity $B/\sqrt{4\pi G\rho}$.  

The Jeans mass of a plasma, subject to magnetic pressure, is then  given by
\begin{equation}
M_{J}=\sqrt{\frac{3}{4\pi G^{3}\rho_{0}}\left(\frac{B^{2}}{4\pi \rho_{0}}+\frac{3}{2}\frac{k_{B}T}{m_{H}\mu}\right)^{3}}  \,\text{ ,}\label{eq:Jeans}
\end{equation}
where $\bar \rho$ is the mean matter density, $m_H$ the mass of a hydrogen atom, $\mu$ the mean molecular weight and $k_B$  the Boltzmann constant.

This expression generalizes previous calculations of the Jeans mass which only considered its limiting cases: $B\rightarrow 0$, the usual Jeans mass (e.g. \citealt{Padmanabhan3}),  or $T\rightarrow 0$, the magnetic Jeans mass (e.g.  \citealt{Tashiro2005}). 

Substituting numerical values and explicitly writing the dependence on cosmological parameters, 
\begin{flalign} 
M_J  = & \left(2.7\times10^4\msun \right)(1+z)^{-3/2} \left(\frac{\Omega_m h^2}{0.133}\right)^{-1/2} \times &
  \\\nonumber &\qquad\qquad
\times \left[T(z)+38\,(1+z)\left(\frac{B_{0}}{2\nG}\right)^2 \left(\frac{\Omega_m h^2}{0.133}\right)^{-1}
\right]^{3/2}\text{ ,} &
\end{flalign}
where $B_{0}$ is the comoving magnetic field and $h$ is the hubble parameter ($H_0=100\, h$ km s$^{-1}$ Mpc$^{-1}$).

It is to be noted that after the recombination era, when the cosmic microwave background was formed, and before the reionization epoch, although the medium was cold, the Universe had a degree of ionization sufficiently high, $x_e\sim 10^{-4}$, to couple magnetic fields to the medium (e.g., \citealt{Padmanabhan3}). This can be understood as follows. In the presence of a magnetic field,  the velocity of ions coupled to the magnetic field (and the relative velocity between ions and neutrals)  becomes almost the Alfven velocity,   which is 1000 times larger than the  thermal velocity of the ions. Because of this,  the time scale of the interaction between ions and neutrals becomes 1000  times shorter and never exceeds the Hubble time before reionization, for the degree of ionization cited above (e.g. \citealt{Tashiro2005}). Thus, a primordial magnetic field was coupled with the medium even before the epoch of reionization.

Having an expression for the Jeans mass that takes into account the primordial magnetic field, we need to know the evolution of the temperature of the gas with redshift. We use the analytic fit of the temperature as a function of redshift that \citet{Kravtsov2004} obtained for the results of \citet{Gnedin2000a} :
\begin{equation}
T(a)=\left\{ \begin{array}{ccl}
                (10^4\text{ K})\left(\frac{1+z_s}{1+z}\right)^\alpha &\text{,}& z>z_s\\
		10^4\text{ K} &\text{,}& z_s \geq  z \geq z_r\\
		(10^4\text{ K})\left(\frac{1+z}{1+z_r}\right)  &\text{,}& z<z_r\\
\end{array}\right.\label{eq:temp}
\end{equation}
where $z>z_s$ is the epoch before the first \HII regions form, $z_r\leq z\leq z_s$ is the epoch of the overlap of multiple \HII regions and  $z<z_r$ is the epoch of complete reionization. \citet{Kravtsov2004} used $z_s\approx 8$ and $z_r\approx 7$ and  $\alpha \approx 6$. In this paper we also use $\alpha=6$.

\begin{figure}
\includegraphics[width=\columnwidth]{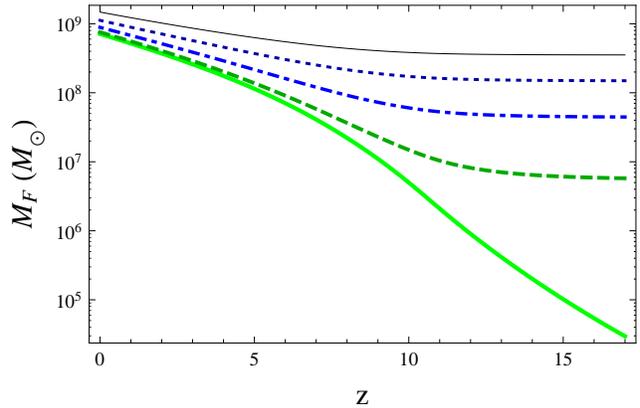}
 \caption{Variation of the filter mass with redshift and magnetic field for $z_s=11$ and $z_r=8$. 
From top to bottom: 
the thick (\textit{green}) curve corresponds to $B_0=0\nG$;
the dashed (\textit{dark-green}) curve, to  $B_0=0.5\nG$;
the dash-dotted (\textit{blue}) curve, to $B_0=1.0\nG$;
the dotted (\textit{dark-blue}) curve, to $B_0=1.5\nG$, and 
the thin (\textit{black}) curve, to $B_0=2\nG$.
}
 \label{fig:filter}
\end{figure}

We use equations (\ref{eq:Jeans}) and (\ref{eq:temp}) in (\ref{eq:filter}) and calculate the effect of a homogeneous magnetic field on the filter mass. The result is shown in figure \ref{fig:filter}, for $z_s=11$ and $z_r=8$. 

\begin{figure}
  \includegraphics[width=\columnwidth]{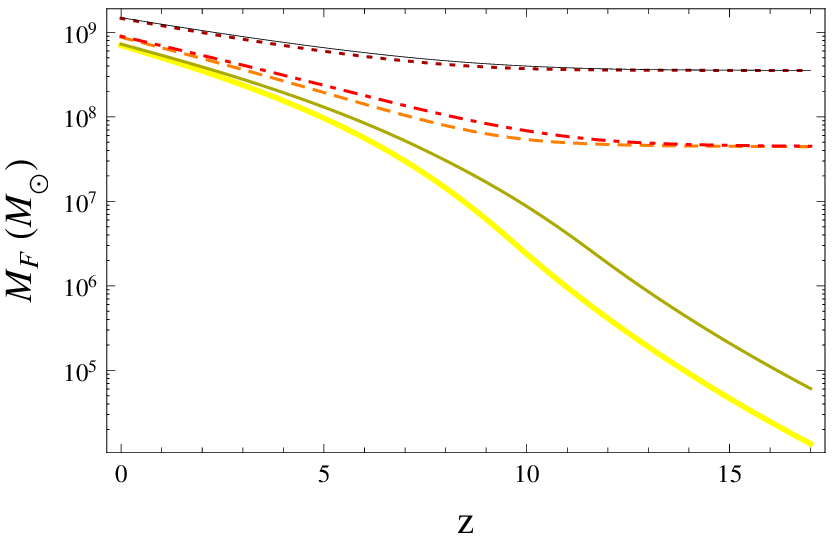}
 \caption{Dependence of the filter mass on the choice of $z_s$ (with fixed $z_r=8$). From top to bottom: 
the thin (\textit{gray}) curve corresponds to $B_0=2\nG$ and $z_s=12$;
the dotted (\textit{dark-red}) curve, to  $B_0=2\nG$ and $z_s=10$;
the dash-dotted (\textit{red}) curve, to $B_0=1\nG$ and $z_s=12$;
the dashed (\textit{orange}) curve, to $B_0=1\nG$ and $z_s=10$;
the continuous (\textit{dark-yellow}) curve, to $B_0=0\nG$ and $z_s=12$, and 
the thick (\textit{yellow}) curve corresponds to $B_0=0\nG$ and $z_s=10$. }
 \label{fig:filteras}	
\end{figure}

\begin{figure}
  \includegraphics[width=\columnwidth]{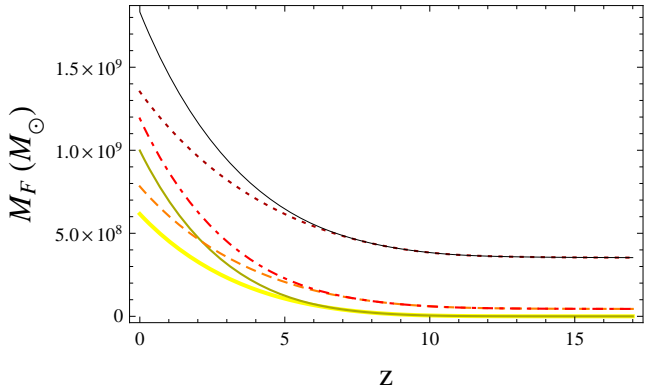}
 \caption{Dependence of the filter mass on the choice of $z_r$ (with fixed $z_s=11$). From top to bottom: 
the thin (\textit{gray}) curve corresponds to $B_0=2\nG$ and $z_r=6$;
the dotted (\textit{dark-red}) curve, to  $B_0=2\nG$ and $z_r=9$;
the dash-dotted (\textit{red}) curve, to $B_0=1\nG$ and $z_r=6$;
the dashed (\textit{orange}) curve, to $B_0=1\nG$ and $z_r=9$;
the continuous (\textit{darkyellow}) curve, to $B_0=0\nG$ and $z_r=6$, and 
the thick (\textit{yellow}) curve corresponds to $B_0=0\nG$ and $z_r=9$. }
 \label{fig:filterar}
\end{figure}

The presence of a homogeneous magnetic field, even for small intensities,  significantly increases the filter mass. We notice that more intense primordial magnetic fields cause a smaller variation of the filter mass with redshift. This last result is expected since, for a homogeneous magnetic field, under the assumption of flux conservation, the magnetic part of equation (\ref{eq:Jeans}) is independent of redshift, i.e.
\begin{equation}
 \left(M_J^{\frac{2}{3}}\right)_{mag}\propto \frac{B^2}{\rho^{4/3}}\propto (1+z)^0 \text{ .}
\end{equation}

The increase in the filter mass leads to less accretion of gas in the smaller haloes, making it more difficult to form small galaxies. The  increase at high redshifts is particularly important since this is the epoch of the formation of galaxies.

We also investigated the dependence of $M_F$ on $z_s$ and $z_r$. Our results are shown in figures  \ref{fig:filteras} and \ref{fig:filterar}, respectively.

From figure \ref{fig:filteras}, with $z_r=8$, it can be seen that changes in $z_s$ (the parameter which marks the begining of the reionization era) only shifts the high-$z$ end of the filter mass curve, without modifying the low-$z$ behaviour.

The effect is, however, much smaller than the increase due to the increase of the primordial magnetic field.

\section{Conclusions}

\label{sec:conclusions}
The Jeans mass was modified in order to account for the presence of a homogeneous primordial magnetic field. We then calculated the implications on the modification of the filter mass. 

The modification leads to a significant increase in the filter mass. We found a change in the behaviour of the filter mass as a function of redshift, which becomes flatter with more intense primordial magnetic fields, as opposed to the tendency of a decline of the filter mass with redshift, which is observed in the absence of magnetic fields.

Our results are shown in figures \ref{fig:filter}--\ref{fig:filterar}. To see the effect of a primordial magnetic field, let us examine its effect at a specific redshift, for example $z=z_s-1$. 

In figure \ref{fig:filter}, for $z_s=11$ and $z_r=8$ we find that the filter mass increases from 
$M_F=5.0\times 10^6\msun$ 
for $B_0=0\nG$ to 
$M_F=6.0\times 10^7\msun$ 
for $B_0=1\nG\,$ and  
$M_F=3.8\times 10^8\msun$ 
for $B_0=2\nG$. 
% ========================
In figure \ref{fig:filteras}, for $z_r=8$ we find that for $z_s=12$ that the filter mass increases from 
$M_F=4.1\times 10^6\msun$ 
 for $B_0=0\nG$ to 
$M_F=5.8\times 10^7\msun$  
for $B_0=1\nG\,$ and  
$M_F=3.8\times 10^8\msun$  
for $B_0=2\nG$. 
% ========================
In figure \ref{fig:filterar}, for $z_s=11$ we find that for $z_r=6$ that the filter mass increases from
$M_F=5.0\times 10^6\msun$ 
 for $B_0=0\nG$ to 
$M_F=6.0\times 10^7\msun$
 for $B_0=1\nG\,$ and  
$M_F=3.8	\times 10^8\msun$
 for $B_0=2\nG$. 

The results of figures \ref{fig:filter}--\ref{fig:filterar} are particularly important for the formation of low mass galaxies. For example, for a reionization epoch $z_s\sim 11$ and $z_r\sim 7$ and a homogeneous primordial magnetic field $B_0=1\nG$, our results indicate that galaxies of total mass $M\sim5 \times10^8\msun$ need to form at redshifts 
$z_F\gtrsim 2.0 $
, and galaxies of total mass $M\sim 10^8\msun$ need to form at  
$z_F\gtrsim 7.7 $.
\section*{Acknowledgments}

L.F.S.R. thanks the Brazilian agency CNPq for financial support (142394/2006-8). 
R.S.S. thanks the Brazilian agency FAPESP for financial support (2009/06770-2). 
R.O.thanks FAPESP (06/56213-9) and the Brazilian agency CNPq (300414/82-0) for partial
support. 
We thank Joshua Frieman and Wayne Hu for helpful comments.
This research has made use of NASA's Astrophysics Data System.

\bibliographystyle{mn2e}
\bibliography{bib}

\bsp

\label{lastpage}
\end{document}